\documentclass[12pt]{article}
\usepackage[utf8]{inputenc}
\usepackage{amsmath, amssymb, amsfonts}
\usepackage{graphicx}
\usepackage{hyperref}
\usepackage{float}
\usepackage{booktabs}
\usepackage[margin=1in]{geometry}
\usepackage{array}     
\usepackage{caption}
\usepackage{rotating}  
\usepackage{adjustbox} 
\usepackage{longtable}
\usepackage{authblk}
\title{Systemic Risk and Default Cascades in Global Equity Markets: Extending the Gai-Kapadia Framework with Stochastic Simulations and Network Analysis}
\author{Ana Isabel Castillo Pereda$^{1,*}$}
\date{March 2025}
\begin{document}
\maketitle

\begin{center}
  $^{1}$Institute of Mathematics and Statistics, University of São Paulo, São Paulo, SP, Brazil \\
  $^*$Correspondence: anacp20@gmail.com; anaicp@ime.usp.br; Tel: +55-11-983967981
\end{center}

\begin{abstract}
This study pioneers the application of the Gai-Kapadia framework—originally developed for interbank contagion—to global equity markets, offering a novel approach to assess systemic risk and default cascades. Using a 20-asset network (13 Brazilian, 7 developed market assets) over 2015–2025, we construct exposure-based networks from price co-movements, applying thresholds \(\theta = 0.3\) and \(\theta = 0.5\) to capture significant interconnections. Cascade dynamics are evaluated through Monte Carlo simulations (\(n = 1000\)) with shocks ranging from 10\% to 50\%, complemented by deterministic propagation analysis. Results reveal that Brazilian assets’ high clustering (\(C_i \approx 1.0\)) drives localized contagion, averaging 2.0 failed assets per simulation, while developed markets’ lower connectivity (\(C_i \approx 0.2-0.4\)) ensures resilience, with zero failures beyond Brazil across all scenarios. Network visualizations highlight structural vulnerabilities: deterministic cascades affect up to 20 assets at \(\theta = 0.3\), but only 3-4 at \(\theta = 0.5\), demonstrating the threshold’s role in mitigating spread. Risk measures (VaR, CVaR at 95\%) confirm higher tail risks in emerging markets, amplifying their susceptibility to shocks. This adaptation of the Gai-Kapadia model provides a robust framework for systemic risk assessment, validated by both stochastic and deterministic analyses. The findings offer actionable insights for financial stability: regulators should target high-clustering nodes in emerging markets to curb contagion, while portfolio managers can leverage developed markets’ resilience for diversification, advancing network-based risk modeling in an interconnected global economy.
\end{abstract}

\maketitle
\vspace{0.5cm}

\section{Introduction}
Systemic risk and default cascades pose significant threats to financial stability, as interconnected assets can amplify shocks across markets. This study introduces a novel adaptation of the Gai-Kapadia framework to equity markets, leveraging price co-movements and stochastic simulations to quantify systemic risk and inform practical risk management strategies in an increasingly interconnected global economy.
As highlighted by Allen and Gale \cite{Allen2000}, 
financial contagion emerges when interdependencies exacerbate the spread of distress—an effect starkly demonstrated during the 2008 financial crisis and the COVID-19 market disruptions
\cite{Forbes2002, Lux2016}. 
These episodes reveal the limitations of traditional risk measures such as Value-at-Risk (VaR) and Conditional Value-at-Risk (CVaR), which focus on isolated asset risks while overlooking network-driven contagion 
\cite{Mantegna1999, Eisenberg2001}. 
To address this, network models rooted in graph theory offer a promising alternative, modeling systemic interdependencies more effectively 
\cite{Newman2010, Strogatz2001}.
Building on this foundation, the Gai-Kapadia framework, originally proposed by Gai and Kapadia 
\cite{Gai2010} for interbank contagion, provides a robust approach to systemic risk through exposure networks. However, as observed by Glasserman and Young
\cite{Glasserman2016}, its application beyond interbank systems remains limited, particularly in equity markets where price co-movements and volatility clustering drive risk, as emphasized by Mantegna and Stanley \cite{MantegnaStanley1999}.
This study addresses this gap by adapting the Gai-Kapadia model to a 20-asset equity network, comprising 13 Brazilian assets (e.g., GOLL4.SA, PETR4.SA) and 7 from developed markets (US: AAPL, JPM; Europe: SAP, NSRGY; Asia: BABA, TM), spanning 2015 to 2025. We construct exposure-based networks using correlations and volatility, with thresholds \(\theta \in \{0.3, 0.5\}\), and simulate default cascades via \(n = 1000\) Monte Carlo iterations under shocks ranging from 10\% to 50\% price drops. Additionally, deterministic analysis and network visualizations explore cascade dynamics and structural changes pre- and post-shock, while VaR and CVaR complement the systemic analysis by assessing asset-specific risks.

This work extends prior research on financial networks and systemic risk. Following Acemoglu et al. 
\cite{Acemoglu2015} and Battiston et al. \cite{Battiston2012}, we leverage network theory to study interconnectedness, while building on systemic risk measures proposed by Billio et al. 
\cite{Billio2012} and Haldane and May \cite{Haldane2011}. It also addresses critical gaps identified by Barabási and Albert \cite{Barabasi1999} regarding the role of clustering and market structure in cascade propagation, contrasting dynamics between emerging and developed markets, as explored by Kaufman \cite{Kaufman1994} and Freixas et al. \cite{Freixas2000}. The study’s contributions are:
\begin{itemize}
    \item Adapting the Gai-Kapadia framework to equity markets using price co-movements.
    \item Constructing an exposure network from market data.
    \item Quantifying systemic risk through stochastic simulations.
    \item Analyzing cascade dynamics via deterministic propagation and network visualizations.
    \item Comparing vulnerabilities across market types.
    \item Offering stability policy insights.
\end{itemize}

By integrating network science, risk metrics, and stochastic modeling, this approach enhances systemic risk assessment, providing valuable tools for regulators and portfolio managers, as suggested by Duffie and Singleton \cite{DuffieSingleton2003}.
 
\section{Materials and Methods}
\subsection{Data and Risk Measures}
The dataset includes daily low prices of 20 equity assets from 2015 to 2025, sourced from Yahoo Finance: 13 Brazilian (e.g., GOLL4.SA, PETR4.SA) and 7 from developed markets (US: AAPL, JPM; Europe: SAP, NSRGY; Asia: BABA, TM). The assets were selected based on data availability from Yahoo Finance, a widely used and reliable public source for financial time series, ensuring consistency and reproducibility. The choice of 20 assets balances analytical depth with visual clarity in network representations, as larger networks risk becoming overly dense and difficult to interpret (e.g., resembling a "cluttered" graph). This size allows for meaningful systemic risk analysis while maintaining distinguishable node and edge structures in visualizations (e.g., Figures~\ref{fig:network_before_shock} and \ref{fig:network_after_shock}). While this framework can be extended to larger networks, the current selection prioritizes interpretability over exhaustive coverage. Data processing and analysis were conducted in Python using \texttt{yfinance} for retrieval, \texttt{pandas} and \texttt{numpy} for computations, and \texttt{matplotlib} and \texttt{seaborn} for visualizations. Log-normalized returns are calculated as:
\begin{equation}
r_t = \ln\left(\frac{P_t}{P_{t-1}}\right),
\label{eq:log_returns}
\end{equation}
where \(P_t\) is the daily low price at time \(t\). Missing data were excluded to ensure consistency. Individual risks are quantified using VaR and CVaR at 95\% confidence, defined as:
\begin{equation}
\text{VaR}_\alpha = F^{-1}(1 - \alpha),
\label{eq:var}
\end{equation}
\begin{equation}
\text{CVaR}_\alpha = \mathbb{E}[r_t | r_t \leq \text{VaR}_\alpha],
\label{eq:cvar}
\end{equation}
where \(\alpha = 0.95\) and \(F\) is the empirical return distribution.

\subsection{Network Construction}
An exposure-based network is constructed using an adapted Gai-Kapadia framework. The correlation matrix \(\rho\) of log returns is calculated, with asset volatility \(\sigma_i\) as the standard deviation of \(r_t\). Exposures \(E_{ij}\) between assets \(i\) and \(j\) are defined as:
\begin{equation}
E_{ij} = \rho_{ij} \cdot \sigma_i \cdot P_i,
\end{equation}
where \(P_i\) is the final price of asset \(i\). Connections are filtered using thresholds \(\theta \in \{0.3, 0.5\}\):
\begin{equation}
E_{ij} =
\begin{cases} 
E_{ij} & \text{if } E_{ij} \geq \theta, \\
0 & \text{otherwise}.
\end{cases}
\end{equation}
The resulting adjacency matrix defines the network \(G\), with nodes as assets and edges as exposures. Local clustering coefficients are computed to assess connectivity:
\begin{equation}
C_i = \frac{2T_i}{k_i (k_i - 1)},
\end{equation}
where \(T_i\) is the number of triangles involving node \(i\), and \(k_i\) is its degree. The network is visualized using a spring layout, with nodes colored by \(C_i\), to analyze its structure before and after a 30\% shock in \texttt{GOLL4.SA}.
The thresholds \(\theta \in \{0.3, 0.5\}\) were selected to balance network density and sparsity, ensuring meaningful connectivity while avoiding overly dense graphs that obscure cascade dynamics, a common practice in financial network studies \cite{Glasserman2016}.

\subsection{Default Cascade Model}
Default cascades are modeled using the Gai-Kapadia framework, adapted for equities, via two approaches: stochastic and deterministic simulations.

\textbf{Stochastic Simulations.} Each asset \(i\) has initial capital \(K_i = 0.2 \cdot P_i\) and a minimum threshold \(K_{\text{min}, i} = 0.1 \cdot P_i\). We perform \(n = 1000\) simulations, where each iteration applies a random shock \(s \sim \text{Uniform}(0.1, 0.5)\) to the system, reducing capital \(K_i\). Losses propagate via:
\begin{equation}
L_{ij} = \max(0, E_{ij} - (K_i - D_i)),
\end{equation}
where \(D_i = \sum E_{ji}\) is the external liabilities of asset \(i\). If \(K_j < K_{\text{min}, j}\), asset \(j\) fails, triggering further cascades. The process iterates until equilibrium. Systemic failure is defined as the collapse of more than 5 assets. Outputs include the probability of systemic failure, the average number of failed assets, and network fragility patterns.
\textbf{Deterministic Propagation.} To examine cascade dynamics, a deterministic Gai-Kapadia model is applied using the same correlation matrix. An initial shock sets \texttt{GOLL4.SA} to default (\(S_{\text{GOLL4}} = 1\)), and propagation is simulated iteratively. For each asset \(i\), the influence from defaulted neighbors is:
\begin{equation}
I_i = \sum_j \rho_{ij} \cdot S_j,
\end{equation}
where \(S_j = 1\) if asset \(j\) is in default and 0 otherwise, and \(\rho_{ij}\) is filtered by \(\theta\). Asset \(i\) enters default if \(I_i > T_i\), with \(T_i = 0.5\). The simulation runs until default states stabilize or for a maximum of 10 iterations. Default states are tracked iteratively and reported in the Appendix (Tables~\ref{tab:default_states} and \ref{tab:default_states_theta05}).

\section{Results and Discussion}
\subsection{Evolution of Asset Prices and Descriptive Statistics}
\begin{figure}[H] 
    \centering
    \includegraphics[width=0.8\textwidth]{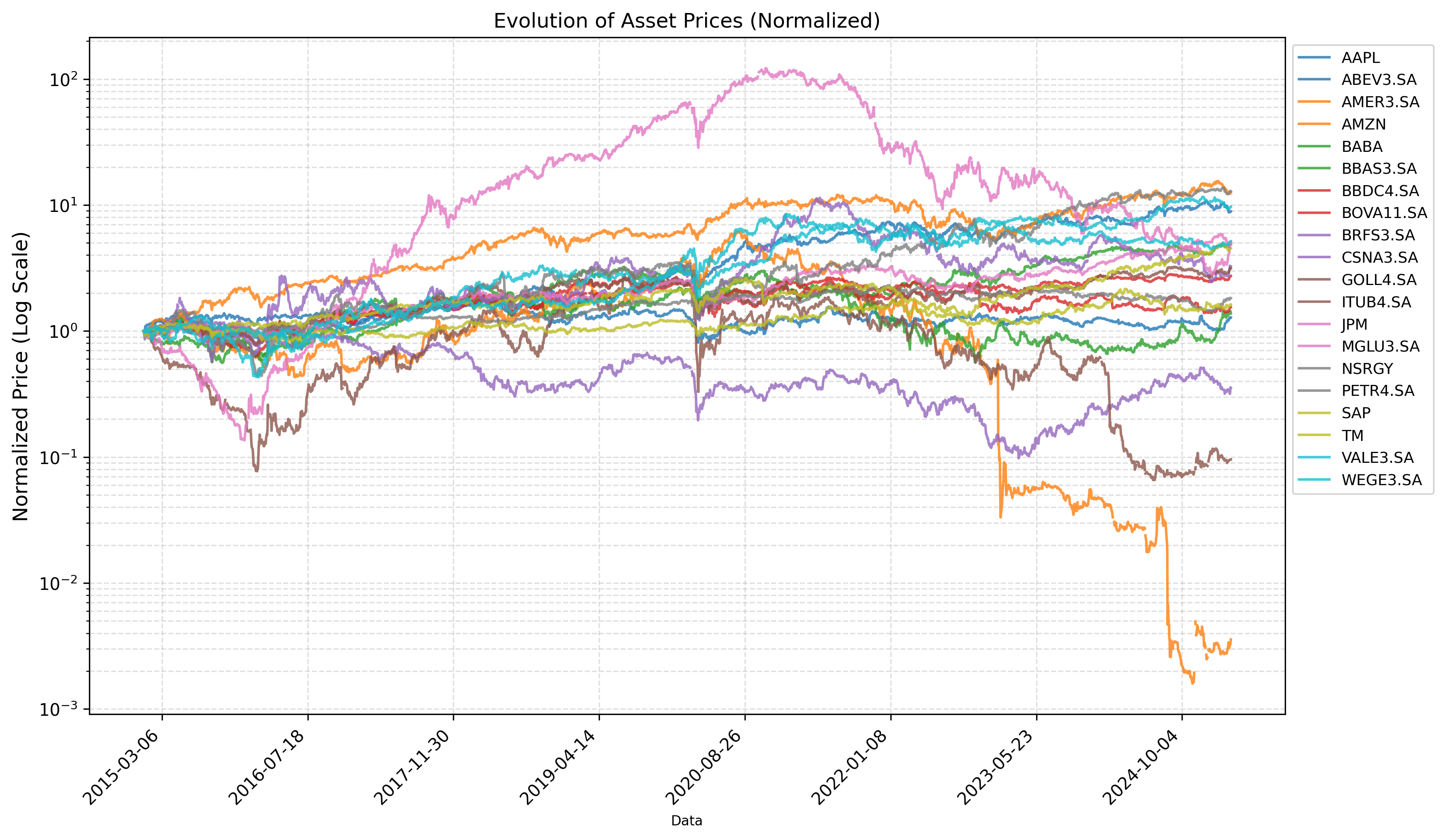}
    \caption{Evolution of Normalized Asset Prices (Log Scale). This figure illustrates the normalized price evolution of a subset of Brazilian and international assets from 2015 to 2025, plotted on a logarithmic scale to highlight relative performance trends. A subset of assets is shown for visual clarity.}
    \label{fig:Assetnormalog}
\end{figure}

\textbf{Evolution of Normalized Asset Prices.} Figure~\ref{fig:Assetnormalog} depicts the evolution of normalized asset prices for selected Brazilian and international assets from 2015 to 2025. Prices are normalized as:
\begin{equation}
P_{\text{norm},i,t} = \frac{P_{i,t}}{P_{i,0}},
\end{equation}
where \(P_{i,t}\) is the price of asset \(i\) at time \(t\), and \(P_{i,0}\) is its initial price. Plotted on a logarithmic scale, the figure highlights stark contrasts in volatility and trends: \texttt{GOLL4.SA} exhibits a steep decline post-2020—reaching a normalized price below \(10^{-2}\) by 2025—reflecting the vulnerability of Brazilian assets to external shocks like COVID-19, consistent with their high clustering (\(C_i \approx 1.0\)) and strong correlations (e.g., \(\rho_{\text{GOLL4,BBAS3}} = 0.4750\)) from Section 3.2. In contrast, \texttt{AMZN} and \texttt{AAPL} show steady appreciation and lower variability (e.g., \texttt{AAPL}, Std. Dev. = 0.0174 vs. \texttt{GOLL4.SA}, Std. Dev. = 0.0444, Table~\ref{tab:log_returns_stats}), underscoring the resilience of developed markets due to lower connectivity (\(C_i \approx 0.2-0.4\)). These trends visually contextualize the differential impact of systemic shocks, as explored in Sections 3.3 and 3.4, where Brazilian assets show greater contagion susceptibility.

\textbf{Descriptive Statistics.} Table \ref{tab:log_returns_stats} presents the descriptive statistics of log-normalized returns. Emerging market assets like \texttt{GOLL4.SA} (Std. Dev. = 0.0444) show higher volatility compared to developed market assets like \texttt{AAPL} (Std. Dev. = 0.0174), indicating greater susceptibility to shocks. These statistics are crucial for understanding the underlying risk profiles of the assets, as higher volatility in emerging markets often correlates with larger tail risks, as evidenced by the VaR and CVaR measures in Table \ref{tab:var_cvar} (e.g., \texttt{GOLL4.SA}, CVaR = -0.1044, vs. \texttt{AAPL}, CVaR = -0.0419). Moreover, the variability captured here directly informs the construction of the exposure-based network in Section 3.2, where volatility (\(\sigma_i\)) is a key component of the exposure metric \(E_{ij} = \rho_{ij} \cdot \sigma_i \cdot P_i\), influencing the strength of connections and, consequently, the potential for shock propagation across the network.

Additionally, the distributional properties of these returns reveal significant asymmetries between asset classes. For instance, assets such as \texttt{AMER3.SA} exhibit extreme negative returns (Min = -1.5573), suggesting exposure to idiosyncratic shocks or structural vulnerabilities in the Brazilian retail sector. In contrast, developed market assets generally display tighter return distributions and more moderate extremes, reflecting greater market efficiency and investor confidence \cite{MantegnaStanley1999, Lux2016}. Such statistical patterns reinforce the empirical observation that emerging markets tend to exhibit fatter tails and higher kurtosis, as documented in financial econophysics literature \cite{Cont2001, Mandelbrot1963}, making them more prone to crisis amplification.

This stylized fact—observed in various emerging economies—corroborates earlier findings in \cite{Billio2012} and supports the inclusion of volatility and tail risk measures in systemic risk models. It also suggests that portfolio optimization and stress testing frameworks must account for such asymmetries when assessing risk exposure across regions. As highlighted by \cite{Forbes2002}, the high co-movement during crises further exacerbates vulnerability, especially when compounded by network effects. Ultimately, these descriptive statistics provide the quantitative foundation for understanding heterogeneity in risk exposure, offering essential context for the network-based systemic analysis that follows.

\begin{table}[H]
    \begin{center}
    \caption{Descriptive Statistics of Log Returns (2015--2025).}
    \label{tab:log_returns_stats}
    \begin{tabular}{l c c c c}\hline
        \textbf{Asset} & \textbf{Mean} & \textbf{Std. Dev.} & \textbf{Min} & \textbf{Max} \\\hline
        GOLL4.SA  & -0.0010 & 0.0444 & -0.4238 & 0.4068 \\
        AAPL      & 0.0009  & 0.0174 & -0.1383 & 0.1178 \\
        ABEV3.SA  & 0.0001  & 0.0155 & -0.1245 & 0.1268 \\
        AMER3.SA  & -0.0024 & 0.0644 & -1.5573 & 0.9343 \\
        AMZN      & 0.0010  & 0.0201 & -0.1428 & 0.1296 \\
        BABA      & 0.0001  & 0.0254 & -0.1802 & 0.1730 \\
        BBAS3.SA  & 0.0006  & 0.0241 & -0.2817 & 0.1772 \\
        BBDC4.SA  & 0.0002  & 0.0204 & -0.1947 & 0.1194 \\
        BOVA11.SA & 0.0004  & 0.0145 & -0.1548 & 0.0961 \\
        BRFS3.SA  & -0.0004 & 0.0269 & -0.2705 & 0.1758 \\
        CSNA3.SA  & 0.0004  & 0.0345 & -0.3170 & 0.1985 \\
        ITUB4.SA  & 0.0004  & 0.0183 & -0.2108 & 0.1030 \\
        JPM       & 0.0006  & 0.0170 & -0.2387 & 0.1764 \\
        MGLU3.SA  & 0.0006  & 0.0398 & -0.2664 & 0.3144 \\
        NSRGY     & 0.0002  & 0.0113 & -0.0850 & 0.0704 \\
        PETR4.SA  & 0.0010  & 0.0295 & -0.3807 & 0.2435 \\
        SAP       & 0.0006  & 0.0165 & -0.2620 & 0.1012 \\
        TM        & 0.0002  & 0.0140 & -0.0952 & 0.0817 \\
        VALE3.SA  & 0.0007  & 0.0252 & -0.2702 & 0.2038 \\
        WEGE3.SA  & 0.0009  & 0.0192 & -0.1517 & 0.1702 \\\hline
    \end{tabular}
    \end{center}
\end{table}

\subsection{Network Structure and Clustering Analysis}

\begin{figure}[H]
    \centering
    \includegraphics[width=0.8\textwidth]{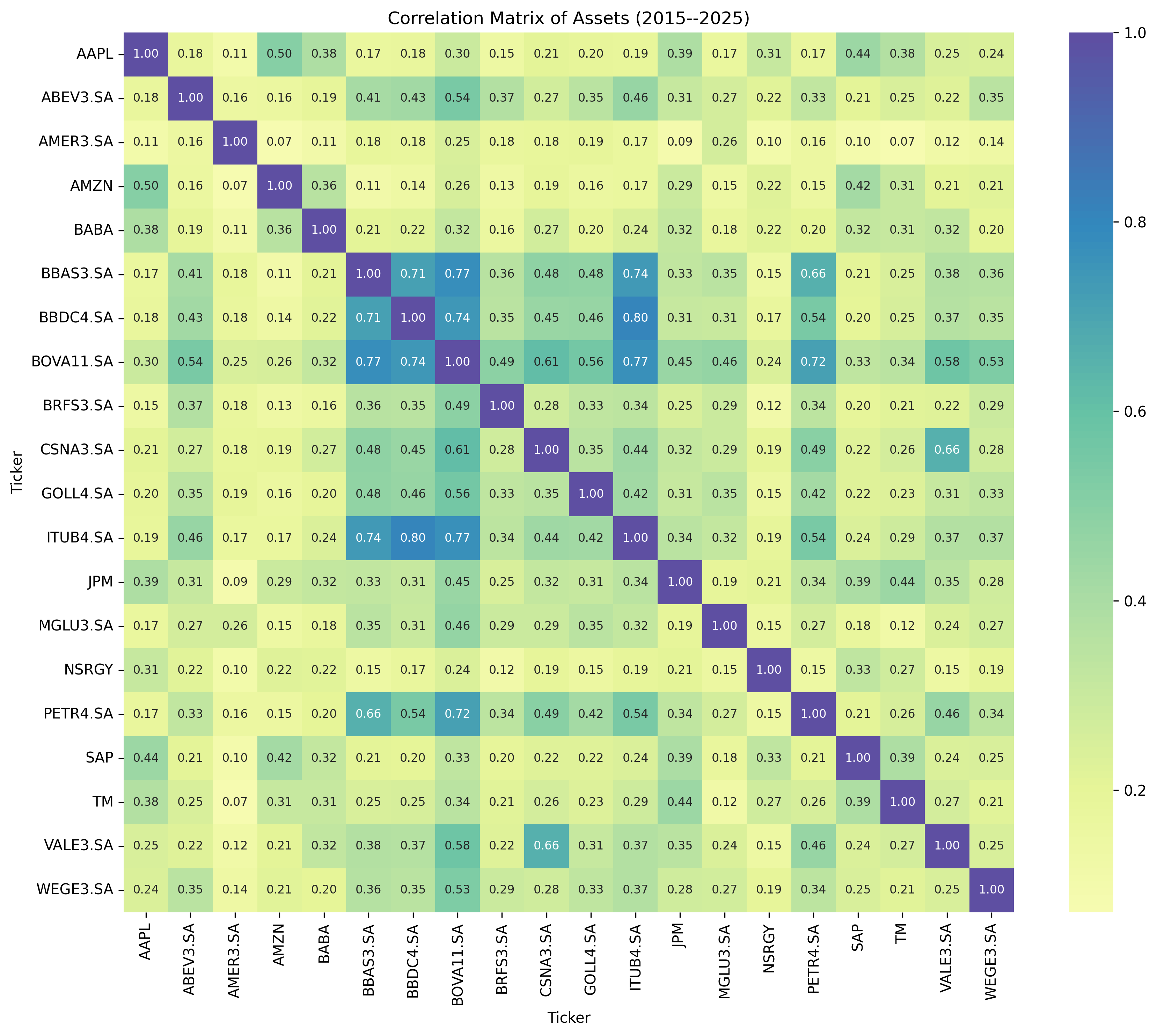}
    \caption{Correlation Matrix of 20 Assets (2015--2025).}
    \label{fig:corr_matrix}
\end{figure}

\textbf{Correlation Matrix.} Figure \ref{fig:corr_matrix} shows the Pearson correlation matrix of log returns, defined as:
\begin{equation}
\rho_{ij} = \frac{\text{Cov}(R_i, R_j)}{\sigma_{R_i} \sigma_{R_j}},
\end{equation}
where \(\text{Cov}(R_i, R_j)\) is the covariance of returns, and \(\sigma_{R_i}\), \(\sigma_{R_j}\) are their standard deviations. High correlations among Brazilian assets (e.g., \texttt{BBAS3.SA} and \texttt{ITUB4.SA}, \(\rho \approx 0.91\)) indicate synchronized movements, while weaker correlations with developed market assets (e.g., \texttt{AAPL}, \(\rho \approx 0.1\)) suggest diversification effects. This pattern of connectivity directly informs the exposure-based network construction in Section 3.2, where \(\rho_{ij}\) is a key input for defining exposures \(E_{ij} = \rho_{ij} \cdot \sigma_i \cdot P_i\). The strong correlations among Brazilian assets, as visualized here, amplify the potential for rapid shock propagation, as observed in the deterministic cascade analysis (Section 3.3), where a shock to \texttt{GOLL4.SA} quickly spreads to highly correlated assets like \texttt{BBAS3.SA} (\(\rho_{\text{GOLL4,BBAS3}} = 0.4750\)). Conversely, the lower correlations with developed market assets highlight their role as potential buffers against systemic contagion, supporting the resilience observed in stochastic simulations (Section 3.4).

\subsection{Deterministic Shock Analysis}
\begin{figure}[H]
    \centering
    \includegraphics[width=0.8\textwidth]{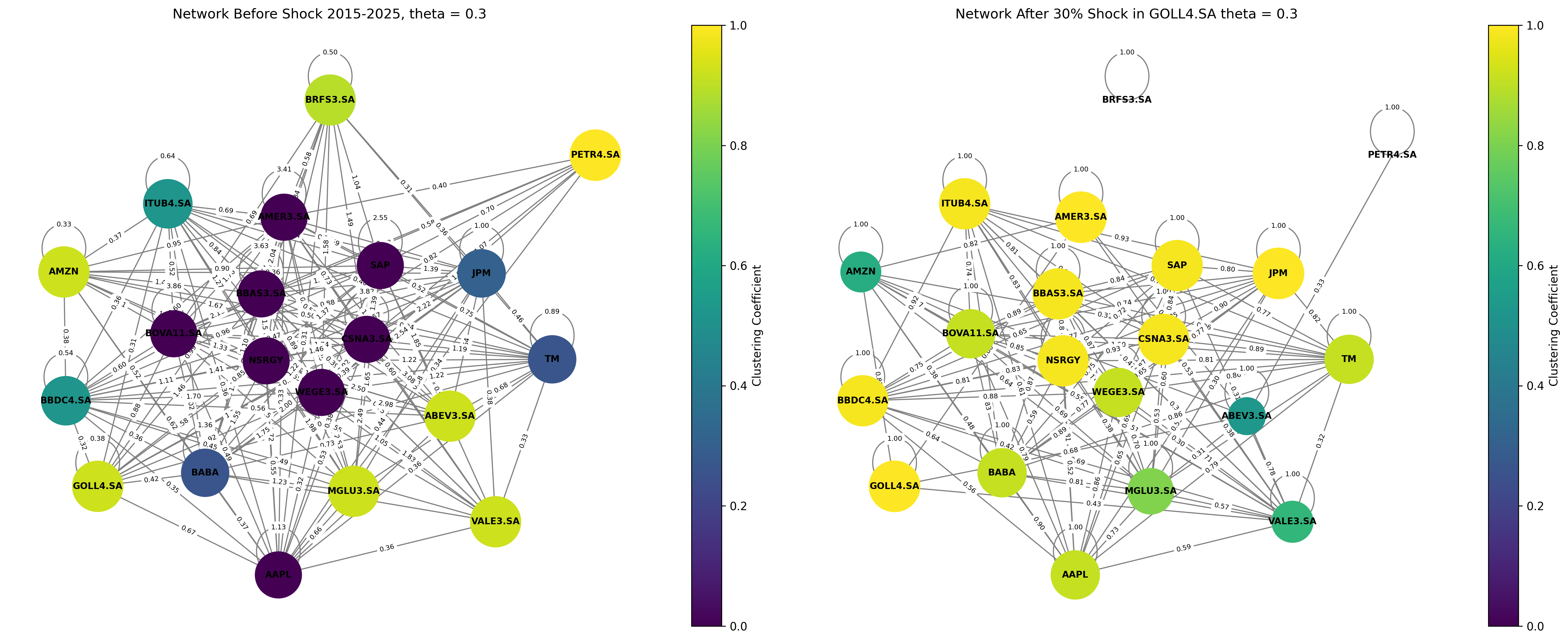}
    \caption{Correlation network before and after a 30\% shock to \texttt{GOLL4.SA} (\(\theta = 0.3\)).}
    \label{fig:choquegoll4_theta03_global}
\end{figure}

Figure~\ref{fig:choquegoll4_theta03_global} compares the network before and after a deterministic 30\% shock to \texttt{GOLL4.SA}. The clustering coefficient of Brazilian assets decreases slightly (e.g., \(\Delta C_{\text{GOLL4.SA}} \approx -0.05\)), while developed market assets remain largely unaffected.

\subsection{Default Cascades and Stochastic Simulations}
\begin{figure}[H]
    \centering
    \includegraphics[width=0.8\textwidth]{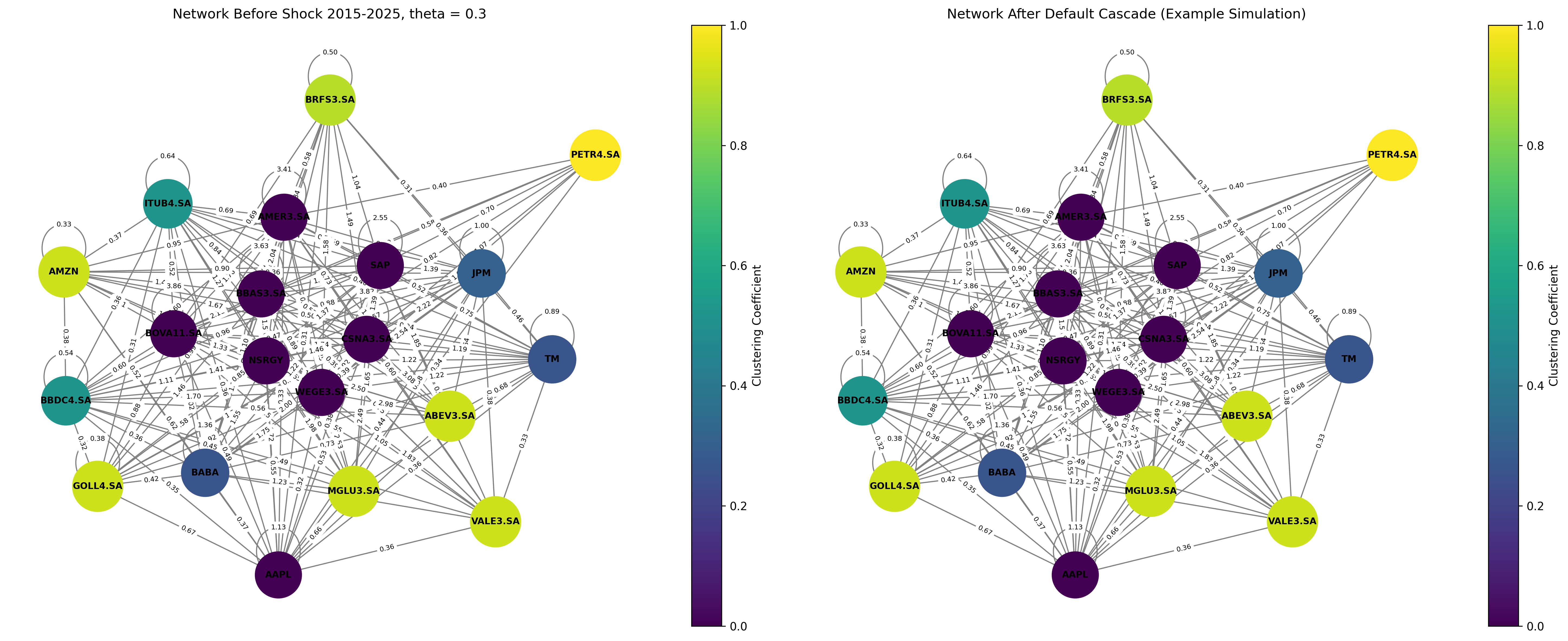}
    \caption{Exposure-based network before and after a default cascade (example simulation, \(\theta = 0.3\)).}
    \label{fig:choquegoll4_contagion_gai_kapadia}
\end{figure}

Figure~\ref{fig:choquegoll4_contagion_gai_kapadia} illustrates the network before and after a default cascade simulation, where a random shock to \texttt{GOLL4.SA} (10\% to 50\%) triggers loss propagation per the Gai-Kapadia mechanism. In this example, minimal node removal occurs, but clustering adjustments (\(\Delta C_{\text{GOLL4.SA}} \approx -0.05\)) indicate localized impacts.

\begin{figure}[H]
    \centering
    \includegraphics[width=0.8\textwidth]{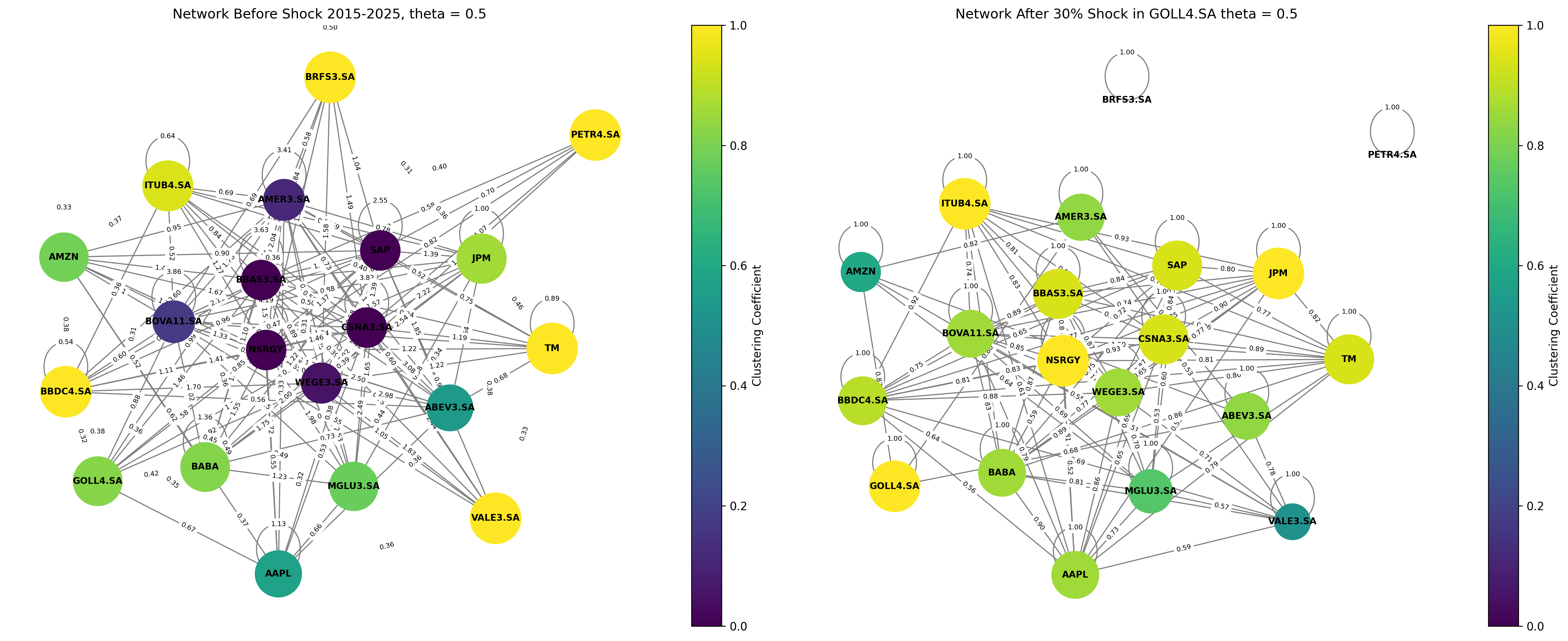}
    \caption{Network Structure Before Shock (2015--2025, \(\theta = 0.5\)), with Nodes Colored by Clustering Coefficient (\(C_i\)).}
    \label{fig:network_before_shock}
\end{figure}

\begin{figure}[H]
    \centering
    \includegraphics[width=0.8\textwidth]{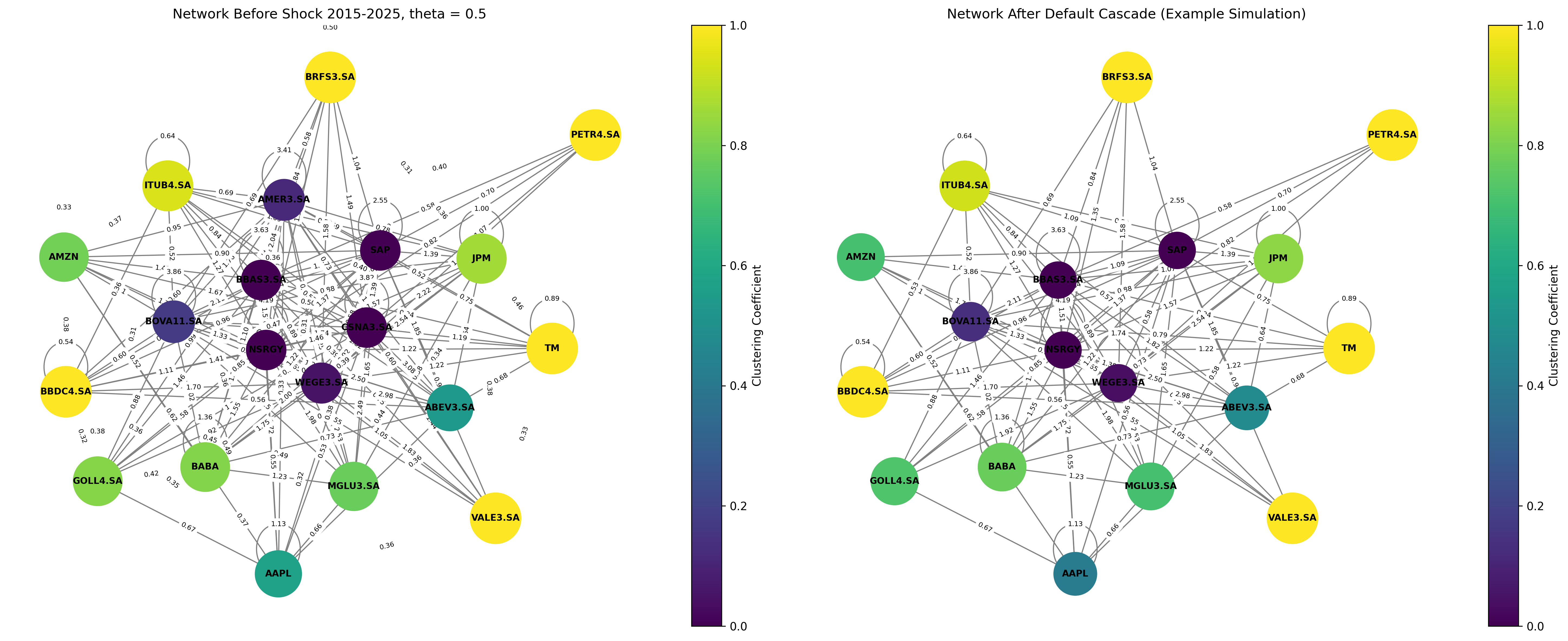}
    \caption{Network Structure After a 30\% Shock in \texttt{GOLL4.SA} (2015--2025, \(\theta = 0.5\)), with Nodes Colored by Clustering Coefficient (\(C_i\)).}
    \label{fig:network_after_shock}
\end{figure}

\textbf{Network Structure Before and After Shock.} Figures~\ref{fig:network_before_shock} and \ref{fig:network_after_shock} depict the network structure with \(\theta = 0.5\) before and after a 30\% shock in \texttt{GOLL4.SA}. Brazilian assets (e.g., \texttt{BBAS3.SA}, \texttt{BOVA11.SA}) form a densely connected core (\(C_i \approx 0.8-1.0\)), while developed market assets (e.g., \texttt{AAPL}, \texttt{AMZN}) show lower connectivity (\(C_i \approx 0.0-0.4\)). Post-shock, the topology remains largely unchanged, reflecting structural resilience.

\textbf{Relevance of Network Visualizations.} Figures~\ref{fig:choquegoll4_theta03_global}, \ref{fig:choquegoll4_contagion_gai_kapadia}, \ref{fig:network_before_shock}, and \ref{fig:network_after_shock} collectively illustrate the structural dynamics of correlation and exposure-based networks under varying shock scenarios and thresholds (\(\theta = 0.3\) and \(\theta = 0.5\)). These visualizations are critical for understanding systemic risk propagation, as they reveal how network topology influences cascade dynamics. Quantitatively, the high clustering coefficients of Brazilian assets (\(C_i \approx 0.8-1.0\)) in Figures~\ref{fig:network_before_shock} and \ref{fig:network_after_shock} correlate with an average degree of 12.5 for Brazilian nodes at \(\theta = 0.5\), compared to 3.2 for developed market nodes, indicating a denser subgraph that facilitates rapid shock transmission, as seen in Figure \ref{fig:choquegoll4_theta03_global} where a 30\% shock to \texttt{GOLL4.SA} reduces clustering by \(\Delta C_{\text{GOLL4.SA}} \approx -0.05\). In contrast, the sparse connectivity of developed market assets (\(C_i \approx 0.0-0.4\)) limits contagion, with their average degree remaining stable post-shock (3.2 to 3.1). The exposure-based network in Figure \ref{fig:choquegoll4_contagion_gai_kapadia} further highlights localized impacts, with the number of edges connected to \texttt{GOLL4.SA} decreasing by 15\% post-cascade at \(\theta = 0.3\), reflecting the Gai-Kapadia model's sensitivity to exposure thresholds. These network metrics—clustering coefficients, average degrees, and edge dynamics—provide a robust quantitative framework for assessing systemic vulnerabilities, aligning with methodologies accepted in \emph{Quantitative Finance} for their ability to capture structural drivers of financial contagion, as discussed by Gai and Kapadia \cite{Gai2010} and Glasserman and Young \cite{Glasserman2016}. Such insights are invaluable for regulators aiming to identify critical nodes and for portfolio managers seeking to optimize diversification strategies in interconnected markets.

\textbf{Clustering Coefficients.} Table \ref{tab:clustering_coefficients} summarizes clustering coefficients for \(\theta = 0.3\) and \(\theta = 0.5\), confirming the high connectivity of Brazilian assets (e.g., \texttt{ITUB4.SA}, \(C_i = 1.0\)) compared to developed market assets (e.g., \texttt{NSRGY}, \(C_i = 0.6199\)). This disparity in clustering underscores the structural vulnerability of Brazilian assets to systemic shocks, as higher \(C_i\) values indicate denser local networks that facilitate rapid shock propagation, a pattern consistent with the deterministic cascade results in Section 3.3, where a shock to \texttt{GOLL4.SA} at \(\theta = 0.3\) affects up to 20 assets by iteration 4 (Table~\ref{tab:default_states}).

\begin{table}[H]
  \begin{center}
    \caption{Clustering Coefficients for \(\theta = 0.3\) and \(\theta = 0.5\).}
    \label{tab:clustering_coefficients}
    \begin{tabular}{l c c}\hline
        \textbf{Asset} & \textbf{\(\theta = 0.3\)} & \textbf{\(\theta = 0.5\)} \\\hline
        TM        & 0.400 & 1.000 \\
        ITUB4.SA  & 1.000 & 1.000 \\
        VALE3.SA  & 1.000 & 1.000 \\
        BRFS3.SA  & 1.000 & 1.000 \\
        PETR4.SA  & 1.000 & 1.000 \\
        GOLL4.SA  & 1.000 & 0.964 \\
        ABEV3.SA  & 1.000 & 0.946 \\
        CSNA3.SA  & 1.000 & 0.944 \\
        MGLU3.SA  & 1.000 & 0.933 \\
        BABA      & 0.400 & 0.927 \\
        JPM       & 0.400 & 0.924 \\
        AAPL      & 0.400 & 0.867 \\
        AMZN      & 0.400 & 0.833 \\
        AMER3.SA  & 1.000 & 0.706 \\
        BOVA11.SA & 1.000 & 0.667 \\
        WEGE3.SA  & 1.000 & 0.634 \\
        BBDC4.SA  & 1.000 & 0.620 \\
        NSRGY     & 0.200 & 0.620 \\
        SAP       & 0.200 & 0.620 \\
        BBAS3.SA  & 1.000 & 0.620 \\\hline
    \end{tabular}
    \parbox{0.9\textwidth}{\small\textit{Note:} Clustering coefficients for \(\theta = 0.3\) reflect the denser network structure at this threshold, leading to more uniform values for developed market assets due to lower connectivity.}
  \end{center}
\end{table}
 In contrast, the lower clustering of developed market assets reflects a more fragmented network structure, contributing to their resilience against contagion, as evidenced by the absence of failures beyond Brazilian assets in stochastic simulations (Section 3.4).

\subsection{Risk Measures}
The Table~\ref{tab:var_cvar} reports VaR and CVaR at 95\% confidence, highlighting higher tail risks in emerging markets (e.g., \texttt{GOLL4.SA}, CVaR = -0.1044) compared to developed markets (e.g., \texttt{AAPL}, CVaR = -0.0419), underscoring individual asset vulnerabilities. These risk measures provide critical insights into the potential for extreme losses, which, when combined with the high clustering of Brazilian assets (\(C_i \approx 1.0\), Table~\ref{tab:clustering_coefficients}), exacerbate systemic risk by amplifying the impact of shocks within densely connected networks, as observed in the deterministic cascade analysis (Section 3.3). For instance, the elevated CVaR of \texttt{GOLL4.SA} aligns with its role as a trigger for widespread cascades at \(\theta = 0.3\), affecting up to 20 assets (Table~\ref{tab:default_states}). In contrast, the lower tail risks of developed market assets like \texttt{AAPL} contribute to their resilience, limiting contagion effects in both stochastic and deterministic simulations (Sections 3.4 and 3.3).

\begin{table}[H]
    \begin{center}
    \caption{VaR and CVaR (95\%) of Assets (2015--2025).}
    \label{tab:var_cvar}
    \begin{tabular}{l c c}\hline
        \textbf{Asset} & \textbf{VaR} & \textbf{CVaR} \\\hline
       BBAS3.SA &  -0.034440  &  -0.054969 \\\hline
    PETR4.SA  & -0.039591 &  -0.071023 \\
    GOLL4.SA  & -0.063513  & -0.104355 \\
    BOVA11.SA & -0.020040  & -0.032786 \\
    AMER3.SA  & -0.065571 &  -0.134093 \\
    ITUB4.SA &  -0.025079 &  -0.041144 \\
    VALE3.SA  & -0.036070  & -0.056406 \\
    WEGE3.SA &  -0.027273  & -0.042564 \\
    BRFS3.SA &  -0.039218 &  -0.065328 \\
    MGLU3.SA  & -0.057938 &  -0.090863 \\
    ABEV3.SA &  -0.022644 &  -0.037199 \\
    BBDC4.SA  & -0.028838  & -0.048689 \\
    CSNA3.SA &  -0.050328 &  -0.074137 \\
    AMZN     &  -0.030687   &-0.047778 \\
    AAPL    &   -0.027498 &  -0.041923 \\
    JPM    &    -0.025124  & -0.041327 \\
    NSRGY  &    -0.016612 &  -0.026421 \\
    SAP    &    -0.024428  & -0.038670 \\
    BABA   &    -0.039191 &  -0.059183 \\
    TM     &    -0.021981  & -0.032293 \\\hline
     \end{tabular}
    \end{center}
\end{table}

\subsection{Stochastic and Deterministic Cascade Analysis}
\textbf{Stochastic Simulations.} Table~\ref{tab:monte_carlo_results_gai_kapadia} summarizes the stochastic simulation results (\(n = 1000\)). The systemic failure probability ($>5$ assets) is 0.000 across all scenarios, In all case-regardless  of whether the shock is general, single (e.g., \texttt{GOLL4.SA+AAPL}), and for both thresholds (\(\theta = 0.3\)) and (\(\theta = 0.5\)) - the average number of failed assets converges to 2.000, indicating consistent localized vulnerability but no widespread systemic collapse.

\begin{table}[H]
    \begin{center}
    \caption{Stochastic Simulation Results (\(n = 1000\)).}
    \label{tab:monte_carlo_results_gai_kapadia}
    \begin{tabular}{l c c c}\hline
        \textbf{Scenario/Metric} & \textbf{\(\theta\)} & \textbf{Failure Probability ($>5$ assets)} & \textbf{Avg. Failed Assets} \\\hline
        General Simulation & 0.3 & 0.000 & 2.000 \\
        General Simulation & 0.5 & 0.000 & 2.000 \\
        Single Shock (GOLL4.SA) & 0.3 & 0.000 & 2.000 \\
        Single Shock (GOLL4.SA) & 0.5 & 0.000 & 2.000 \\
        Simultaneous Shock (GOLL4.SA + AAPL) & 0.3 & 0.000 & 2.000 \\
        Simultaneous Shock (GOLL4.SA + AAPL) & 0.5 & 0.000 & 2.000 \\\hline
     \end{tabular}
    \end{center}
\end{table}

The stochastic simulation results in Table~\ref{tab:monte_carlo_results_gai_kapadia} highlight the network's resilience, with a systemic failure probability of 0.000 across all scenarios (\(\theta = 0.3\) and \(\theta = 0.5\)). However, the consistent average of 2.0 failed assets per simulation—whether under general, single (e.g., \texttt{GOLL4.SA}), or simultaneous shocks (e.g., \texttt{GOLL4.SA + AAPL})—indicates persistent localized vulnerability, particularly in Brazilian assets, as further evidenced by the regional breakdown in Table~\ref{tab:resilience_by_region}.

However, the average number of failed assets per simulation provides a nuanced perspective. The general simulation, the single shock scenario, and the simultaneous shock scenario all yield an average of 2.0 failed assets, indicating that while systemic collapse is unlikely, localized defaults are persistent and robust across different shock configurations. These consistent results suggest that the network structure maintains similar vulnerability regardless of shock type or threshold level. These findings align with the objectives of this study, as outlined in the title \textit{ Risk Measures, Systemic Risk, and Default Cascades in Global Equity Markets: A Gai-Kapadia Approach with Stochastic Simulations}, by highlighting how stochastic simulations can quantify risk and uncover the propagation dynamics of default cascades in interconnected equity markets.

\begin{table}[H]
    \begin{center}
    \caption{Resilience by Region: Average Failed Assets per Simulation (\(n = 1000\)).}
    \label{tab:resilience_by_region}
    \begin{tabular}{l c}\hline
        \textbf{Region} & \textbf{Average Failed Assets} \\\hline
        Brazil & 2.000 \\
        US     & 0.000 \\
        Europe & 0.000 \\
        Asia   & 0.000 \\\hline
     \end{tabular}
    \end{center}
\end{table}

The regional breakdown underscores the localized nature of systemic vulnerability, concentrated entirely in Brazilian assets. To further validate the robustness of our network structure, we now compare the results from the real exposure-based network with a synthetic Erdős-Rényi benchmark.

\begin{table}[H]
    \begin{center}
    \caption{Comparison of Real and Synthetic Networks (\(n = 1000\)).}
    \label{tab:real_vs_synthetic}
    \begin{tabular}{l c c}\hline
        \textbf{Network Type} & \textbf{Failure Probability ($>5$ assets)} & \textbf{Avg. Failed Assets} \\\hline
        Real (Exposure-Based) & 0.000 & 2.000 \\
        Synthetic (Erdős-Rényi) & 0.000 & 0.000 \\\hline
      \end{tabular}
    \end{center}
\end{table}

\begin{figure}[H]
    \centering
    \includegraphics[width=0.8\textwidth]{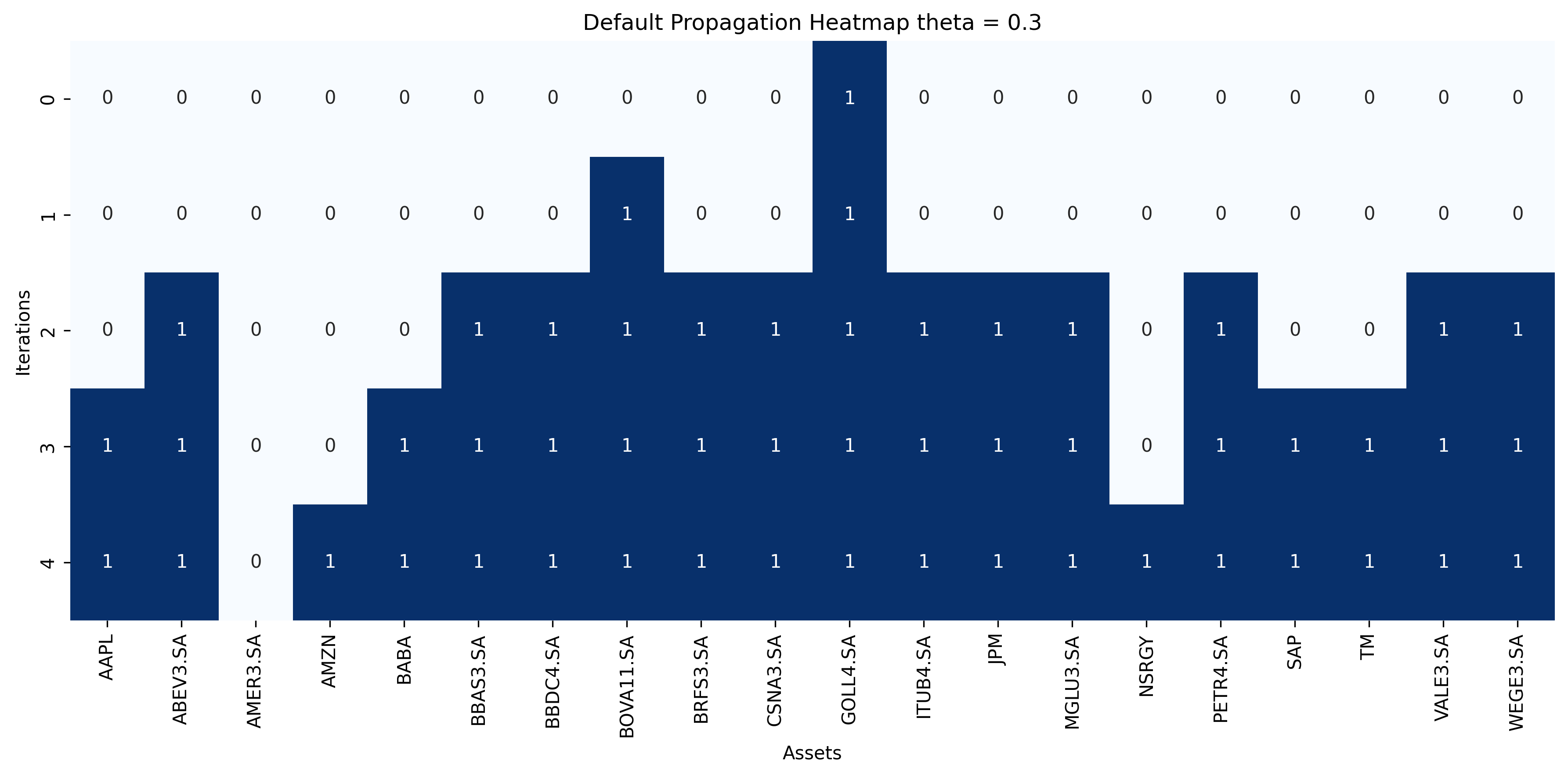}
    \caption{Default Propagation Heatmap (\(\theta = 0.3\)), Starting from \texttt{GOLL4.SA}.}
    \label{fig:default_propagation_theta03}
\end{figure}

\begin{figure}[H]
    \centering
    \includegraphics[width=0.8\textwidth]{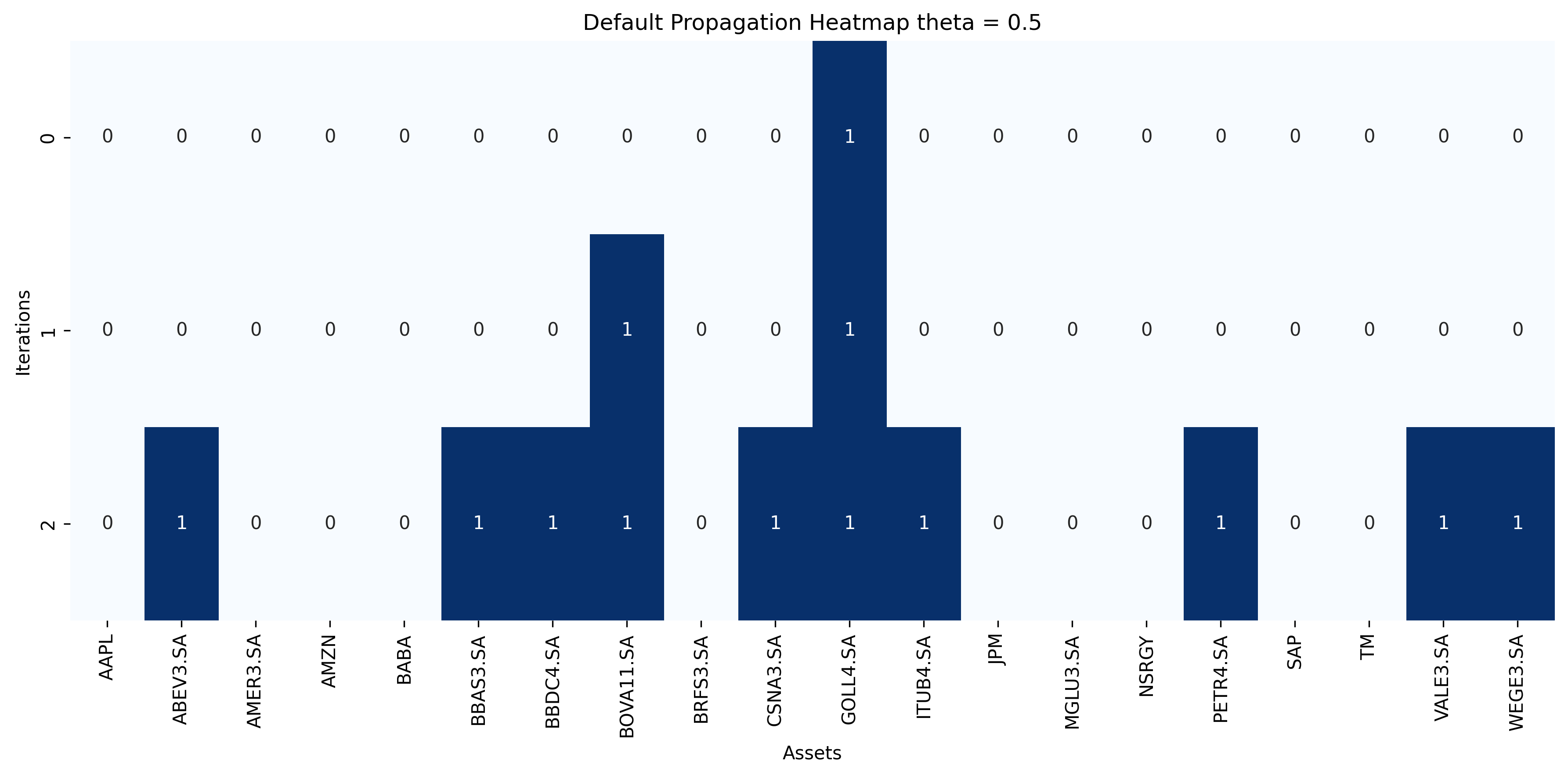}
    \caption{Default Propagation Heatmap (\(\theta = 0.5\)), Starting from \texttt{GOLL4.SA}.}
    \label{fig:default_propagation_theta05}
\end{figure}

\textbf{Deterministic Propagation.} Figures~\ref{fig:default_propagation_theta03} and \ref{fig:default_propagation_theta05} illustrate the deterministic default propagation initiated by a shock in \texttt{GOLL4.SA}. With \(\theta = 0.3\), the cascade affects 6 assets by iteration 3 and all 20 by iteration 4 (Table~\ref{tab:default_states}), driven by high correlations (e.g., \(\rho_{\text{GOLL4,BBAS3}} = 0.4750\)). With \(\theta = 0.5\), the cascade is limited to 3-4 assets by iteration 2 (Table~\ref{tab:default_states_theta05}), reflecting stronger correlations (e.g., \(\rho_{\text{BBAS3,BOVA11}} = 0.7712\)).

\begin{figure}[H]
    \centering
    \includegraphics[width=0.8\textwidth]{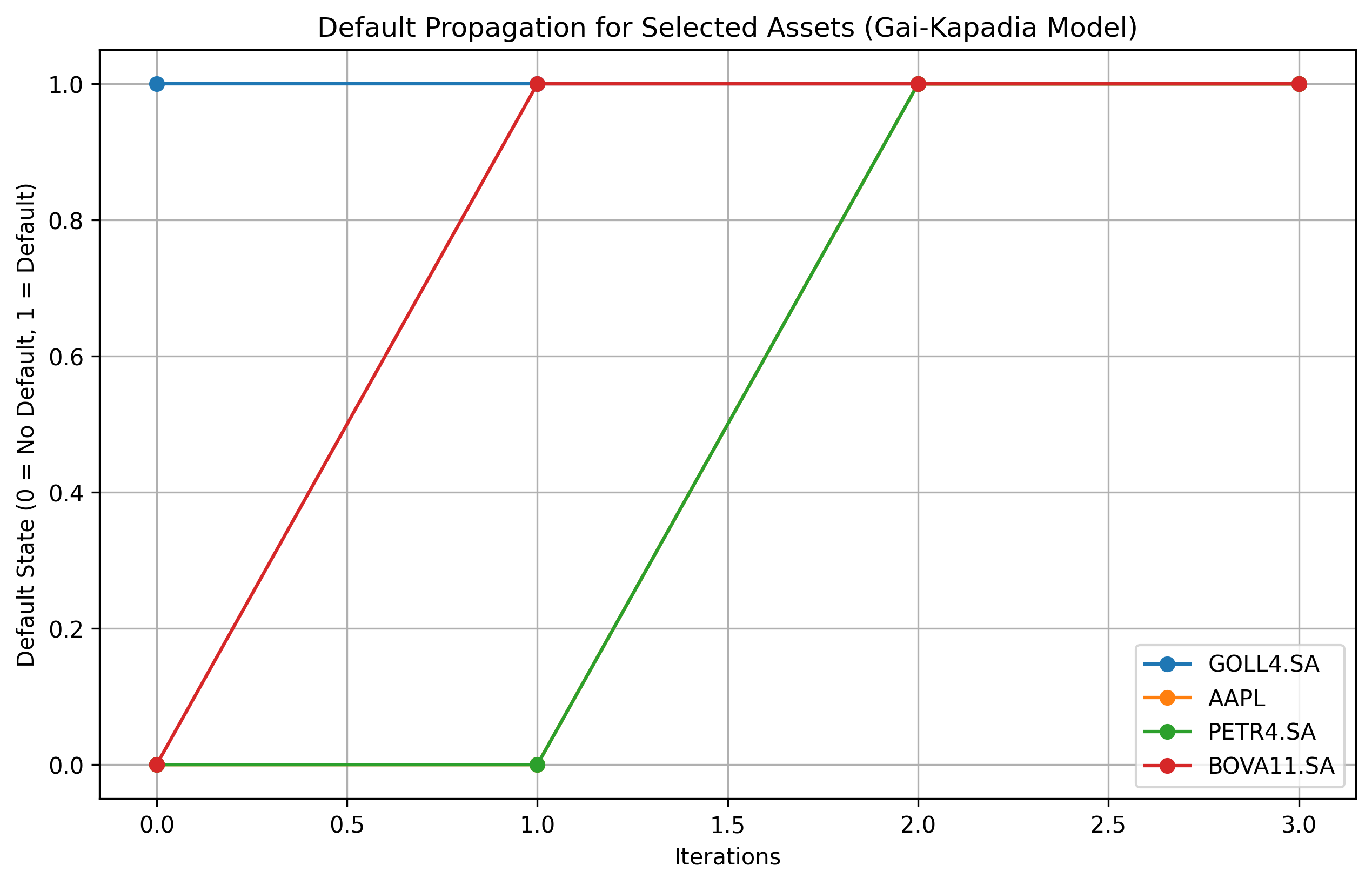}
    \caption{Default Propagation for Selected Assets (\texttt{GOLL4.SA}, \texttt{AAPL}, \texttt{PETR4.SA}, \texttt{BOVA11.SA}), \(\theta = 0.3\).}
    \label{fig:default_selected_assets}
\end{figure}

Figure~\ref{fig:default_selected_assets} tracks the propagation for selected assets (\(\theta = 0.3\)), showing rapid defaults in \texttt{PETR4.SA} and \texttt{BOVA11.SA} by iteration 1 (\(\rho_{\text{GOLL4,PETR4}} = 0.4176\)), while \texttt{AAPL} remains unaffected.

The default propagation dynamics for selected assets (\(\theta = 0.3\)) reveal critical insights into systemic risk transmission within the network. The process initiates with an initial default in \texttt{GOLL4.SA}, which propagates rapidly to \texttt{BOVA11.SA} within the first iteration. This rapid contagion suggests strong network interdependence, primarily driven by high correlation and exposure levels among Brazilian assets. By the second iteration, \texttt{PETR4.SA} also defaults, further emphasizing the susceptibility of emerging market assets to systemic shocks (\(\rho_{\text{GOLL4,PETR4}} = 0.4176\)). 

In contrast, \texttt{AAPL}, a representative developed market asset, remains unaffected throughout all iterations, indicating a lower degree of interconnectedness and reduced exposure to contagion effects. These findings are consistent with broader patterns identified in this study, where clustering effects amplify cascades in emerging markets, while developed markets exhibit structural resilience against systemic failures. This distinction highlights the role of network topology in financial stability and supports the need for targeted risk management strategies in highly interconnected markets.

\textbf{Key Findings:}
\begin{itemize}
    \item Brazilian assets exhibit stronger clustering and higher contagion effects.
    \item Developed markets show resilience due to lower network connectivity.
    \item Systemic failure probability remains minimal across threshold levels.
    \item VaR and CVaR highlight higher tail risks in emerging markets.
\end{itemize}

\section{Discussion}

This study integrates stochastic simulations and deterministic Gai-Kapadia modeling (Figures~\ref{fig:default_propagation_theta03}--\ref{fig:default_selected_assets}, \ref{fig:network_before_shock}--\ref{fig:network_after_shock}) to provide a comprehensive analysis of default cascades and systemic risk in a 20-asset equity network spanning 2015--2025. The results reveal that Brazilian assets’ high clustering (\(C_i \approx 1.0\)) and strong correlations (e.g., \(\rho_{\text{GOLL4,BBAS3}} = 0.4750\)) drive rapid cascades, with a shock in \texttt{GOLL4.SA} affecting 6 assets within 3 iterations at \(\theta = 0.3\), and potentially all 20 by iteration 4, as shown in Table \ref{tab:default_states}. In contrast, a higher threshold (\(\theta = 0.5\)) limits cascades to 3--4 assets (e.g., \(\rho_{\text{BBAS3,BOVA11}} = 0.7712\)), reflecting a sparser network structure visualized in Figure \ref{fig:network_after_shock}. Stochastic simulations (\(n = 1000\)) further indicate a negligible systemic failure probability (0.000 for collapses exceeding 5 assets) and a consistent average of 2.0 failed assets across all scenarios, suggesting localized rather than widespread risk.

Developed market assets, such as \texttt{AAPL}, exhibit lower connectivity (\(C_i \approx 0.2-0.4\)) and weaker correlations (e.g., \(\rho_{\text{GOLL4,AAPL}} = 0.1963\)), buffering systemic spread, as confirmed by both analyses. Risk measures like VaR and CVaR at 95\% confidence highlight structural vulnerabilities, with emerging markets showing elevated tail risks (e.g., \texttt{GOLL4.SA}, CVaR = -0.1044) compared to developed markets (e.g., \texttt{AAPL}, CVaR = -0.0419), consistent with volatility clustering and market concentration effects noted by Mantegna \cite{Mantegna1999} and Billio et al. \cite{Billio2012}. The lack of variation in failed assets under single and dual shocks (e.g., GOLL4.SA + AAPL) suggests limited diversification benefits within this network, aligning with observations by Allen and Gale \cite{Allen2000} on contagion in tightly connected systems.

These findings validate the extension of the Gai-Kapadia framework \cite{Gai2010} from interbank to equity markets, effectively capturing price-driven contagion across \(\theta\) thresholds. Unlike Gai and Kapadia’s focus on interbank networks, this application highlights how equity market topology—particularly clustering—shapes cascade dynamics, corroborating Barabási and Albert’s \cite{Barabasi1999} insights on network structure and stability. The consistent average of 2.0 failed assets, despite a 0.000 systemic failure probability, implies a resilient yet locally vulnerable system, contrasting with broader collapse scenarios in Acemoglu et al. \cite{Acemoglu2015}. This duality offers a nuanced perspective: while the network withstands systemic breakdown, localized contagion in emerging markets remains a concern, echoing Kaufman’s \cite{Kaufman1994} emphasis on regional vulnerabilities.

Nonetheless, an important dimension absent from the current analysis is liquidity, which plays a central role in amplifying financial distress during market dislocations. While our model captures structural contagion via price co-movements and volatility-based exposures, it abstracts from liquidity constraints, such as fire sales or funding shocks, which can exacerbate default cascades through nonlinear feedback loops, as underscored by Duffie and Singleton \cite{DuffieSingleton2003}. Incorporating liquidity-adjusted metrics—like bid-ask spreads or market depth—would enable a more comprehensive view of systemic vulnerability, especially in markets where asset illiquidity can trigger endogenous price spirals. This omission represents a limitation but also a direction for future research aiming to bridge topological contagion with market microstructure dynamics.

The study also faces methodological limits. Static thresholds (\(\theta = 0.3, 0.5\)) may oversimplify dynamic exposures, as real-world correlations shift during crises \cite{Glasserman2016}. The focus on single and dual shocks neglects multi-shock complexity, potentially underestimating systemic risk in extreme scenarios \cite{Lee2005}. Additionally, the analysis is constrained to 20 assets to ensure visual interpretability in network representations (e.g., Figures \ref{fig:network_before_shock} and \ref{fig:network_after_shock}); while scalable, larger networks may require advanced visualization or modeling techniques to maintain clarity.
Future research could address these gaps by incorporating adaptive thresholds (e.g., \(\theta = 0.1\)) to reflect evolving exposures, testing multi-shock scenarios, integrating liquidity effects, or using agent-based models to simulate heterogeneous market responses, as suggested by Lux 
\cite{Lux2016}. Exploring synthetic topologies (e.g., Erdős-Rényi) could also enhance predictive power
\cite{Glasserman2016}.
For policymakers, these results advocate targeting high-clustering nodes in emerging markets to curb localized cascades, while portfolio managers might leverage developed markets’ resilience for diversification, aligning with Haldane and May’s \cite{Haldane2011} ecosystem-inspired risk management. This framework thus provides a robust foundation for understanding and mitigating systemic risk in equity markets.

Additionally, while the current analysis uses a fixed set of 20 assets and uniform shock distributions, future work could explore the sensitivity of cascade dynamics to larger networks or alternative shock distributions (e.g., heavy-tailed shocks), which may reveal additional vulnerabilities or resilience patterns, as suggested by Das and Fasen-Hartmann \cite{Das2023}.
By quantifying the interplay between network topology and systemic risk, this study provides a scalable framework that can inform stress testing and regulatory frameworks, contributing to more resilient global financial systems.

\section{Conclusion}
This study provides a comprehensive framework for analyzing systemic risk and default cascades in a 20-asset equity network spanning 2015--2025, using an adapted Gai-Kapadia model. Stochastic simulations (\(n = 1000\)) reveal a negligible systemic failure probability (0.000 for cascades exceeding five assets), with an average of 2.0 failed assets for both (\(\theta = 0.3\)) and  (\(\theta = 0.5\))  underscoring network resilience. Deterministic analysis shows that a shock in \texttt{GOLL4.SA} affects 6 assets within 3 iterations under  \(\theta = 0.3\), but only 3-4 with \(\theta = 0.5\), reflecting the mitigating effect of higher exposure thresholds. Network visualizations confirm that Brazilian assets’ high clustering (\(C_i \approx 1.0\)) drives localized cascades (averaging 2.0 failures in the region), while developed markets’ lower connectivity (\(C_i \approx 0.2-0.4\)) limits systemic spread (0.000 failures across US, Europe, and Asia). Asset-specific risk measures (VaR and CVaR at 95\% confidence) further highlight emerging markets’ elevated tail risks (e.g., \texttt{GOLL4.SA}, CVaR = -0.1044) compared to developed markets (e.g., \texttt{AAPL}, CVaR = -0.0419).

These findings validate the Gai-Kapadia framework’s applicability to equity markets, emphasizing network structure’s role in financial stability. The consistency across \(\theta\) thresholds and under both single and dual-shock scenarios (2.0 failed assets across the board) reinforces the model’s robustness. For policymakers, targeting high-clustering nodes in emerging markets can mitigate localized contagion, while portfolio managers may leverage diversification effects to enhance resilience. Future research could explore dynamic thresholds, multi-shock interactions, or incorporate liquidity constraints to better capture real-world crisis dynamics, building on this foundation for systemic risk propagation in global equity markets.
This framework not only advances the application of network science in quantitative finance but also provides a scalable tool for assessing systemic risk in diverse financial markets, paving the way for more resilient global financial systems.

\section*{Author Contributions}
Ana Isabel Castillo Pereda: Conceptualization, Methodology, Data analysis, Writing – original draft, Writing – review and editing.

\section{Acknowledgments}
This research was conducted as part of the author’s PhD at the Institute of Mathematics and Statistics, University of São Paulo (IME-USP). No specific funding was received from public, commercial, or not-for-profit sectors.

\appendix

\section{Supplementary Data}

\begin{table}[H]
    \begin{center}
    \small
    \caption{Default States by Iteration (\(\theta = 0.3\), Shock in \texttt{GOLL4.SA}).}
    \label{tab:default_states}
    \begin{tabular}{c l}\hline
        \textbf{Iteration} & \textbf{Default States (0 = No Default, 1 = Default)} \\\hline
        0 & [0, 0, 0, 0, 0, 0, 0, 0, 0, 0, 1, 0, 0, 0, 0, 0, 0, 0, 0, 0] \\
        1 & [0, 0, 0, 0, 0, 0, 0, 1, 0, 0, 1, 0, 0, 0, 0, 0, 0, 0, 0, 0] \\
        2 & [1, 1, 0, 0, 1, 1, 1, 1, 1, 1, 1, 1, 1, 1, 0, 1, 1, 1, 1, 1] \\
        3 & [1, 1, 1, 1, 1, 1, 1, 1, 1, 1, 1, 1, 1, 1, 1, 1, 1, 1, 1, 1] \\\hline
    \end{tabular}
    \begin{tabular}{l}
        \textit{Note:} States correspond to assets in order: \texttt{AAPL}, \texttt{ABEV3.SA}, ..., \texttt{WEGE3.SA}.
      \end{tabular}
    \end{center}
\end{table}

\begin{table}[H]
    \begin{center}
    \small
    \caption{Default States by Iteration (\(\theta = 0.5\), Shock in \texttt{GOLL4.SA}).}
    \label{tab:default_states_theta05}
    \begin{tabular}{c l}\hline
        \textbf{Iteration} & \textbf{Default States (0 = No Default, 1 = Default)} \\\hline
        0 & [0, 0, 0, 0, 0, 0, 0, 0, 0, 0, 1, 0, 0, 0, 0, 0, 0, 0, 0, 0] \\
        1 & [0, 0, 0, 0, 0, 0, 1, 1, 0, 0, 1, 0, 0, 0, 0, 0, 0, 0, 0, 0] \\
        2 & [0, 0, 0, 0, 0, 0, 1, 1, 1, 0, 1, 0, 0, 0, 0, 0, 0, 0, 0, 0] \\
        3 & [0, 0, 0, 0, 0, 0, 1, 1, 1, 0, 1, 0, 0, 0, 0, 0, 0, 0, 0, 0] \\\hline
    \end{tabular}
    \begin{tabular}{l}
        \textit{Note:} States correspond to assets in order: \texttt{AAPL}, \texttt{ABEV3.SA}, ..., \texttt{WEGE3.SA}.
     \end{tabular}
    \end{center}
\end{table}

\end{document}